%% file: main.tex
\documentclass[10pt,conference]{IEEEtran}

\usepackage{courier}
\usepackage[ruled,vlined]{algorithm2e}
\usepackage{graphicx}
\usepackage{paralist}
\usepackage{tabularx}
\usepackage{balance}
\usepackage{multirow}
\usepackage{multicol}
\usepackage{pbox}
\usepackage{mathtools}
\usepackage[TABBOTCAP]{subfigure}
\usepackage{algorithmic}
\usepackage{paralist}
\usepackage{tabularx}
\usepackage{url}
\usepackage{balance}
\usepackage{multirow}
\usepackage{multicol}
\usepackage{amsmath}
\usepackage{setspace}
\usepackage[table,xcdraw]{xcolor}
\usepackage{makecell}
\usepackage{verbatim}
\usepackage{float}
\usepackage[normalem]{ulem}
\usepackage{xspace}
\usepackage{amssymb}
\usepackage{ifthen}
\usepackage{pifont}
\usepackage{tikz}

\usepackage{hyperref}

\usepackage{booktabs}
\usepackage{multirow}
\usepackage{color,soul}
\usepackage{wrapfig}

\include{macro}
\newcommand{\ie}{\textit{i.e.,}\xspace}

\newcommand{\etal}{\textit{et al.}\xspace}
\newcommand{\secref}[1]{Sec.~\ref{#1}\xspace}

\newcommand{\figref}[1]{Fig.~\ref{#1}\xspace}

\newcommand{\tabref}[1]{Table~\ref{#1}\xspace}

\newcommand{\SecureReqNet}{\textit{SecureReqNet}\xspace}

\begin{document}
%

\title{Learning to Identify Security-Related \\Issues Using Convolutional Neural Networks}


\author{\IEEEauthorblockN{David N. Palacio, Daniel McCrystal, Kevin Moran, \\ Carlos Bernal-C\'ardenas, Denys Poshyvanyk}
\IEEEauthorblockA{Department of Computer Science\\
College of William \& Mary\\
Williamsburg, VA\\
Email: \{danaderp, dmrccr, kpmoran, cebernal, denys\}@cs.wm.edu}
\and
\IEEEauthorblockN{Chris Shenefiel}
\IEEEauthorblockA{Principal Engineer\\
Advanced Security Research Group\\
Cisco Systems\\
Email: cshenefi@cisco.com}
}



\maketitle

\begin{abstract}

Software security is becoming a high priority for both large companies and start-ups alike due to the increasing potential for harm that vulnerabilities and breaches carry with them. However, attaining robust security assurance while delivering features requires a precarious balancing act in the context of agile development practices. One path forward to help aid development teams in securing their software products is through the design and development of security-focused automation. Ergo, we present a novel approach, called \SecureReqNet, for automatically identifying whether issues in software issue tracking systems describe security-related content. Our approach consists of a two-phase neural net architecture that operates purely on the natural language descriptions of issues. The first phase of our approach learns high dimensional word embeddings from hundreds of thousands of vulnerability descriptions listed in the CVE database and issue descriptions extracted from open source projects. The second phase then utilizes the semantic ontology represented by these embeddings to train a convolutional neural network capable of predicting whether a given issue is security-related. We evaluated \SecureReqNet by applying it to identify security-related issues from a dataset of thousands of issues mined from popular projects on GitLab and GitHub. In addition, we also applied our approach to identify security-related requirements from a commercial software project developed by a major telecommunication company. Our preliminary results are encouraging, with \SecureReqNet achieving an accuracy of 96\% on open source issues and 71.6\% on industrial requirements.
\end{abstract}

\input{intro}
\input{approach}
\input{eval}
\input{results}
\input{related-work}
\input{conclusion}

%
\IEEEpeerreviewmaketitle


\section*{Acknowledgement}

This work was supported in part by a grant from the Cisco Advanced Security Research Group. Any opinions, findings, and conclusions expressed herein are the authors and do not necessarily reflect those of our sponsors.

\balance
\bibliographystyle{abbrv}
\bibliography{references.bib}

\end{document}

%% file: intro.tex
\section{Introduction}
\label{sec:intro}

Software has pervaded nearly every facet of modern society. From banking, to art, to entertainment, the benefits that the general public has enjoyed from the automation and ease of use afforded by software applications is unprecedented. However, the pervasive nature of modern software systems is a double-edged sword, as more security- and privacy- sensitive data is being managed by software than ever before. Thus, the potential negative impacts of software vulnerabilities or breaches is higher than ever. This can be readily seen in the fallout from high-profile data breaches of companies such as Experian~\cite{equifax}. Furthermore, past studies related to security requirements have observed that development teams often do not focus on security early in the development process~\cite{MelladoCSI'10} and reconciling the implementation of security patches and fixes with general maintenance related to bugs is a challenging task as software evolves~\cite{Li:CCS'17}. As the implications of software security and safety have grown, so has the need for practitioners to take proactive measures to ensure the integrity of their software systems during the development process.

Security procedures have become deeply integrated into agile software development practices principles~\cite{agile-manifesto} due to the constraints these practices impose~\cite{Siponen:CSS'05}. For instance, companies like Cisco Systems, Inc. and Microsoft Corp. have been generating and adopting standards such as the Secure Development Lifecycle (SDL)~\cite{cisco,microsoft} to effectively integrate security practices into the Agile development process. However, in order to enable these development teams to implement proactive software security practices, new tools are required that automate repetitive or intellectually-intensive tasks that would otherwise be delayed or abandoned during typical software development and maintenance cycles. In this paper, we aim to enable such automation by designing and evaluating an approach that is capable of automatically differentiating between security-related (SR) and non-security related (non-SR) issues. This binary classification is performed in issue tracking systems using as input only the Natural Language (NL) contents of issue descriptions. 

Software teams often use issue tracking systems to record requirements, organize their work, and manage bug reporting. As such, issues often contain information related to the security requirements of an associated software project. We posit that an effective technique for automating the identification of issues that contain SR information can aid developers in at least the following four ways: (i) when issues are used to specify NL software requirements, automated identification of SR issues could help developers more easily identify and prioritize overlooked security-critical requirements; (ii) issues that contain incoming security-sensitive bug reports can be readily identified and more effectively prioritized and resolved; (iii) the contents of automatically identified security issues could be used to refine and focus future automated techniques for traceability or feature location towards higher effectiveness on security-critical software artifacts; (iv) identified SR issues could be more effectively assigned to engineers with skills and experience in  security-critical software development.

Given the potentially positive impact that an effective automated security issue identification approach might have on developers, in this paper we design and implement a novel technique for this problem based upon Natural Language processing (NLP) techniques and Convolutional Neural Networks (CNNs). Our key insight is that existing large corpora of open source issues and security vulnerabilities can be mined and used to learn rich semantic word embeddings. These embeddings can then be used as a feature representation to train a shallow/deep neural network to distinguish between SR and non-SR related issues. We implemented this approach, which we call \SecureReqNet, using several variations of CNN architectures specialized for NLP tasks. We applied our approach to learn embeddings from a dataset combining text extracted from (i) 52,908 CVE entries, (ii) 53,214 issues extracted from the GitLab CE issue tracker and a set of prominent and diverse projects mined from GitHub, and (iii) 10,000 general articles mined from Wikipedia. We then trained the different variations of CNN architectures on a subset of these CVEs, and both security and non-security critical software issues. Finally, we evaluate  the end-to-end performance of our different variations of SecureReqNet on a set of 1,032 unseen issues. Additionally, we worked with Cisco Systems, Inc. to evaluate \SecureReqNet on a set of 69 requirements from a closed-source commercial software project. The results of our evaluation are promising, with the most effective variation of SecureReqNet achieving an accuracy of 96\% on the set of open source issues and 71.6\% on the set of industrial requirements. 

In summary, this paper makes the following contributions:

\begin{itemize}
	\item{The design and implementation of a new technique, called \SecureReqNet for automatically identifying SR issues in issue trackers using a combination of unsupervised and supervised deep learning techniques.}
	\item{A comprehensive evaluation of our approach on both open source issues and closed source requirements extracted from a commercial project developed by a major telecommunications company.}
	\item{An online appendix~\cite{appendix} which includes a freely available, open source implementation of our approach, pre-trained models and weights to facilitate reproducibility, and all the data\footnote{We are not able to release the commercial dataset from our industrial partner due to NDA restrictions.} used to train and evaluate our approach.}
\end{itemize}

%% file: approach.tex
\section{Automatically Identifying Security Related Issues}
\label{sec:approach}

We propose \SecureReqNet, a binary classifier (SR from non-SR issues) that combines neural word embeddings with adapted CNN architectures. We implemented four architecture variations of \SecureReqNet via adaptions to the shape of the tensors, the size of the kernels in the convolutional layers, and the number of neurons in the fully connected layers.

\begin{figure}[t]

\centering
\vspace{-0.3cm}
\includegraphics[width=0.67\columnwidth]{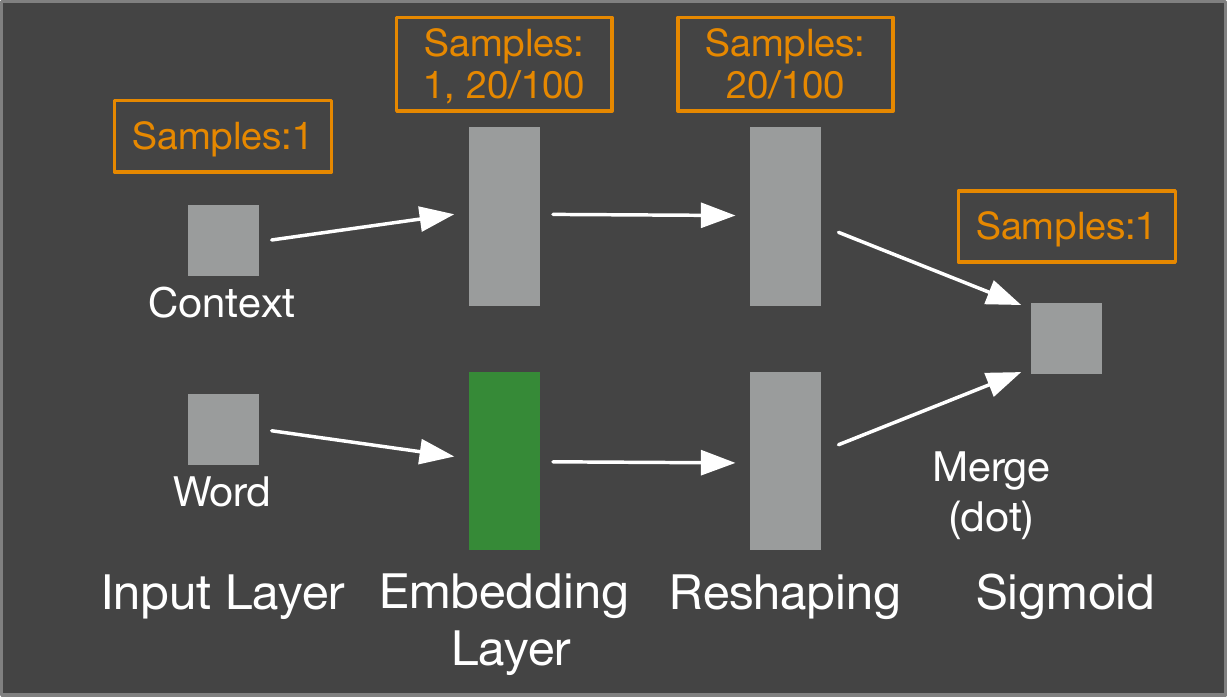}
\vspace{-0.2cm}
\caption{The unsupervised Word2Vec network receives a word and its context (skip-gram), each of which are mapped to a high dimensional vector. The output layer is a softmax function that predicts the probability of the context of the word. The network is trained using the mean squared error. Then, the word vectors are extracted from the embedding layer (green box).}
\label{fig:embedding}
\end{figure}

\subsection{Learning Semantic Word Embeddings}
\label{subsec:embeddings}

In order to obtain a rich semantic feature space to support training our CNN supervised models, our approach learns neural word embeddings using Word2Vec~\cite{Mikolov:arXiV:'13}. Recent work~\cite{Han:ICSME'18,Chen:ASE'16,Xu:ASE'16} has illustrated the effectiveness of neural language models learned using the Word2Vec family of models~\cite{Mikolov:arXiV:'13}.

The Word2Vec group of models uses a shallow neural network trained to predict the current word given surrounding context (\ie the continuous bag-of-words CBOW model) or the surrounding context given the current word (\ie the skip-gram model). \figref{fig:embedding} depicts the unsupervised network employed to vectorize the input samples (\ie preprocessed issues) in \SecureReqNet. The result is a rich, high dimensional vector space wherein each unique word in a training corpus is assigned a vector. These word vectors are concatenated to assemble a matrix. The rows indicate words in an issue description and the columns the embedding dimension. Therefore, each issue has an appropriate matrix that we use as an input sample to \SecureReqNet.

\vspace{-0.1cm}
\subsection{Adapting CNNs for NLP to Predict Security-related Issues}
\label{subsec:cnn-arch}
\vspace{-0.1cm}

In this subsection we describe the four variations of \SecureReqNet's CNN architecture. We provide complete implementations of these architectures in our online appendix~\cite{appendix}.

\textbf{\SecureReqNet (shallow)} is based on the best architecture achieved by Han \etal \cite{Han:ICSME'18}. This architecture establishes one convolution layer with three kernels of different shapes. Each kernel corresponds to either a 1-gram, 3-gram, or 5-gram. The kernel reduces the input matrix into a feature map. Then the feature map is reduced into one cell through max pooling. The generated cells are then flattened and converted into a vector. Each kernel generates one vector. The final tensor is a merged vector from the three vectors produced by the initial kernels. Unlike Han \etal's SVM multi-class output layer \cite{Han:ICSME'18}, we instead utilize a binary classifier through a softmax layer. 

\textbf{\SecureReqNet (deep)} is an extension of \SecureReqNet (shallow) that includes an extra convolutional layer with an extra max pooling and flattening function. 

\textbf{\textit{Alex-SecureReqNet} (deep)} is based on the proposed architecture by Krizhevsky \etal (AlexNet) \cite{Krizhevsky:NIPS13}, where 5 convolutional layers extract features and 3 fully connected layers reduce the dimensionality. We employ the standard AlexNet but adapt the final softmax layer for our binary classification. 

\begin{figure}[t]
\centering
\vspace{-0.3cm}
\includegraphics[width=\columnwidth]{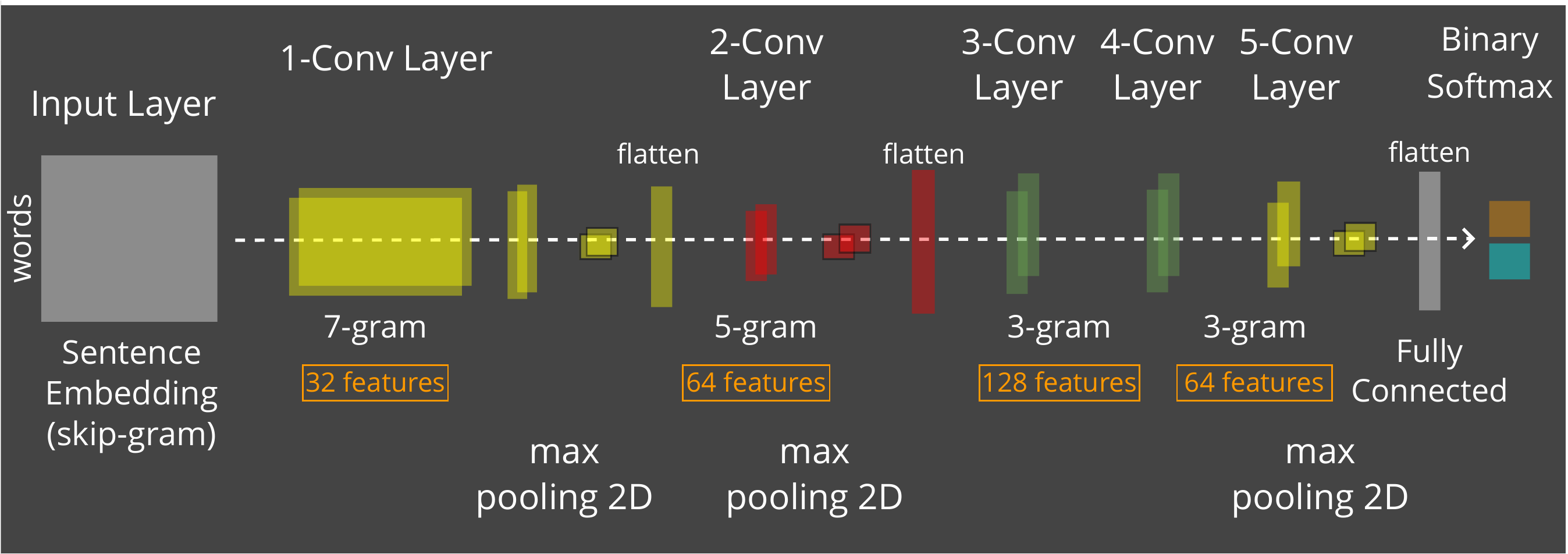}
\vspace{-0.6cm}
\caption{$\alpha$-SecureReqNet CNN Architecture.}
\label{fig:architecture}
\vspace{0.1cm}
\end{figure}

\textbf{\textit{$\alpha$-SecureReqNet} (deep)} modifies \textit{Alex-SecureReqNet} at the convolutional layers. The modification consists of adopting the n-gram kernel strategy from Han \etal \cite{Han:ICSME'18} which should combine the learning capacity of AlexNet with the textual context awareness provided by Han \etal's approach. \figref{fig:architecture} provides an illustration of this architecture. 
The first convolutional layer has a kernel of 7-gram to reduce the input embedding matrix into 32 vector feature maps. Then, it applies a max pooling and a flattening function to derive a one column matrix. The second convolutional layer has a kernel of 5-gram followed by a max pooling and flattening function, which merges other 64 features into another column matrix.  The third, fourth, and fifth convolutional layers are very similar to the original layers in AlexNet; however, these layers model kernels of 3-gram and produce 128 and 64 features for the third and fourth layers respectively. Three fully connected layers go after the last convolutional layer to reduce the dimensionality and control overfitting by means of three dropout units.

%% file: eval.tex
\section{Empirical Evaluation}
\label{sec:eval}

To evaluate \SecureReqNet, we perform an empirical evaluation with two major \textit{goals}: (i) determine which CNN architecture performs the best in terms of identifying SR issues, and (ii) evaluate the effectiveness of our best performing model on datasets from different domains. The \textit{quality focus} of our study is our approach's effectiveness in terms of its ability to accurately distinguish between SR and non-SR issues. To help guide our evaluation, we formulate the following two research questions (RQs):

\begin{itemize}
	\item{\textbf{RQ$_1$:}\textit{ Which \SecureReqNet CNN architecture performs best in terms of accurately identifying SR issues?}} 
\item{\textbf{RQ$_2$:}\textit{ How well does our best CNN model perform when identifying SR issues from both open source and industrial issue trackers?}} 
\end{itemize}

\subsection{Study Context}
\label{subsec:study-context}

The \textit{context} of our empirical study includes the four datasets (\textit{Embedding}, \textit{Training}, \textit{Validation}, \& \textit{Test}) illustrated in \tabref{tab:datasets}.  These datasets are comprised of textual documents from six different sources: 

\noindent \textit{\textbf{CVE Database:}} Our CVE Dataset was derived by crawling the National Vulnerability Database (NVD) and extracting the vulnerability description for each CVE entry. In total, we extracted over 100,000 CVE descriptions, however, in order to construct a dataset balanced equally between SR and non-SR text, we randomly sampled 52,908 CVE descriptions. 

\noindent \textit{\textbf{GitLab Issues:}} To obtain a large set of diverse issues extracted from the issue trackers of a high-quality open source project we crawled the issue tracker of the GitLab Community Edition (CE) project\footnote{\url{https://gitlab.com/gitlab-org/gitlab-ce/issues}}. This project contains open source components of the GitLab suite of developer tools (used by millions) with an issue tracker that includes a sophisticated labeling system. To extract SR issues, we crawled this issue tracker and extracted issue descriptions containing ``security'' label. To extract non-SR issues we extracted entries without the ``security'' label and manually verified the non-SR nature of the descriptions by randomly sampling of 10\% of the issues.

\input{tab/datasets.tex}

\noindent \textit{\textbf{GitHub Issues:}} Given the limited number of SR GitLab issues that we were able to extract, we also crawled the issue trackers of the most popular projects on GitHub (according to number of stars) and extracted issues with the ``security'' tag in order to derive a larger and more diverse dataset. Again, we randomly crawled non-SR issues and performed a random sampling to ensure the validity of the non-SR issues. We provide the full list of GitHub projects in our online appendix~\cite{appendix}.

\noindent \textit{\textbf{Wikipedia Articles:}} If we trained our neural embeddings on \textit{only} highly specialized software text extracted from issues, we risk our model not learning more generalized word contexts that could help differentiate between SR and non-SR issues. Thus, we randomly crawled and extracted the text from 10,000 Wikipedia articles in order to bolster the generalizablility of our learned neural word embeddings. 

\noindent \textit{\textbf{Industrial Dataset:}} As part of our empirical evaluation we aim to determine whether our approach can effectively identify SRs in a set of commercial requirements extracted from a software project developed by a large telecommunications company. This dataset (labeled as ``Industry'' in \tabref{tab:datasets}) consists of a set of users stories, 24 of which are security- or privacy- sensitive, extracted from a project under active development at Cisco Systems, Inc.. Given the more informal language utilized in these requirements, we view this dataset as a more challenging domain for our approach to generalize to, and a preliminary evaluation of its potential for commercial applicability.

The \textit{Embedding Dataset} was used during the unsupervised training of the Word2Vec skip-gram model in order to derive rich embeddings for both SR and non-SR requirements. To construct the \textit{Training}, \textit{Validation}, and \textit{Test} datasets, we split the \textit{Embedding Dataset} according to a 70\%/20\%/10\% (training/validation/test) split. Note that the training set utilizes CVE descriptions as a proxy for SR issues, as these descriptions are inherently SR and thus should provide sufficient discriminatory context for our CNN model to learn.  The data split we utilized was performed with careful consideration for the type of data our approach operates upon and thus adheres to two important properties. First, we include \textit{only} issues extracted from open source issue trackers (and industrial user stories) in our test set. We did \textit{not} include CVE as examples of SR issues. This is because CVEs tend to have more security-specific language than SR issues, and we aim to determine if our approach trained on CVEs can generalize to the more general (or colloquial) language typically used in software issues. Second, the software issues used in our training and testing sets are temporal in nature. Thus, in a typical use case of training our model, it would not be possible to train on issues from the future to identify SR issues from the past. In fact this could provide the model with ``clairvoyant'' knowledge which could unfairly inflate our model's performance. Therefore, we ensure that the issues included in the testing set are both unseen by the model, and originally created temporally \textit{after} all issues in the training and validation sets to ensure a realistic experimental environment. All text was preprocessed by converting all text to lower-case, removing stop words numbers and punctuation, and performing snowball stemming.

\subsection{Evaluation Methodology}
\label{subsec:eval-methods}

For our empirical evaluation, we trained the four architectures described in \secref{subsec:cnn-arch} with the \textit{Training Dataset} described previously. Each input document was vectorized into a 100 dimensional embedding using the neural embedding approach described in \secref{subsec:embeddings}. The networks were trained using a dropout rate of $0.2$. The networks were also subject to an early stopping mechanism, which monitored the validation loss for 2,000 epochs with a patience of 100 epochs. During training, we saved the optimal model weights according to the early-stopping mechanism in addition to the final weights. We tested the optimally trained configuration of each of the four architectures on the \textit{Test Dataset} described earlier. Given that we aim to differentiate the open source project issues from our set of industrial requirements, we report results for these two sources of information separately. To measure the effectiveness of our four architectures we report the accuracy (given our binary classification task), test set loss, and Area Under the Curve (AUC) computed via an ROC analysis.

%% file: tab/datasets.tex
\newcolumntype{C}[1]{>{\centering\let\newline\\\arraybackslash\hspace{0pt}}m{#1}}
\newcolumntype{?}{!{\vrule width 1pt}}
\begin{table}[t]
\vspace{-0.4cm}
\caption{Datasets utilized in the empirical evaluation. security-related requirements (\textbf{SR}) and non security-related requirements (\textbf{Non-SR}).}
\label{tab:datasets}
\footnotesize
\begin{tabular}{C{2.14cm}|C{1.4cm}|C{1.05cm}|C{1.25cm}|C{0.87cm}}
\Xhline{2\arrayrulewidth} & \textbf{Embedding Dataset} & \textbf{Training Dataset} & \textbf{Validation Dataset} & \textbf{Testing Dataset} \\ \cline{2-5} 
\multirow{-2}{*}{\textbf{Dataset Source}} & \multicolumn{4}{c}{\textbf{Number of Documents}}                                       \\ \Xhline{2\arrayrulewidth}
CVE Database                                                      & 52,908                      & 37,036                     & 10,582                       & \textbf{-}                        \\ \hline
GitLab (SR)                                                & 578                        & 405                       & 116                         & 58                       \\ \hline
GitLab (Non-SR)                                            & 578                        & 405                       & 116                         & 58                       \\ \hline
GitHub (SR)                                                & 4,575                       & 3,203                      & 915                         & 458                      \\ \hline
GitHub (Non-SR)                                            & 47,483                      & 33,238                     & 9,497                        & 458                      \\ \hline
Wikipedia                                                         & 10,000                      & \textbf{-}                         & \textbf{-}                           & \textbf{-}                        \\ \Xhline{2\arrayrulewidth}
Industry (SR)                                                         &   \textbf{-}                    & \textbf{-}                         & \textbf{-}                           & 24                       \\ \hline
Industry (Non-SR)                                                         & \textbf{-}                     & \textbf{-}                         & \textbf{-}                           & 45                        \\
\Xhline{2\arrayrulewidth}
\end{tabular}
\end{table}

%% file: results.tex
\section{Empirical Results}
\label{sec:results}

\begin{figure}[t]
\centering
\vspace{-0.3cm}
\hspace*{-0.3cm}
\includegraphics[width=1.05\columnwidth]{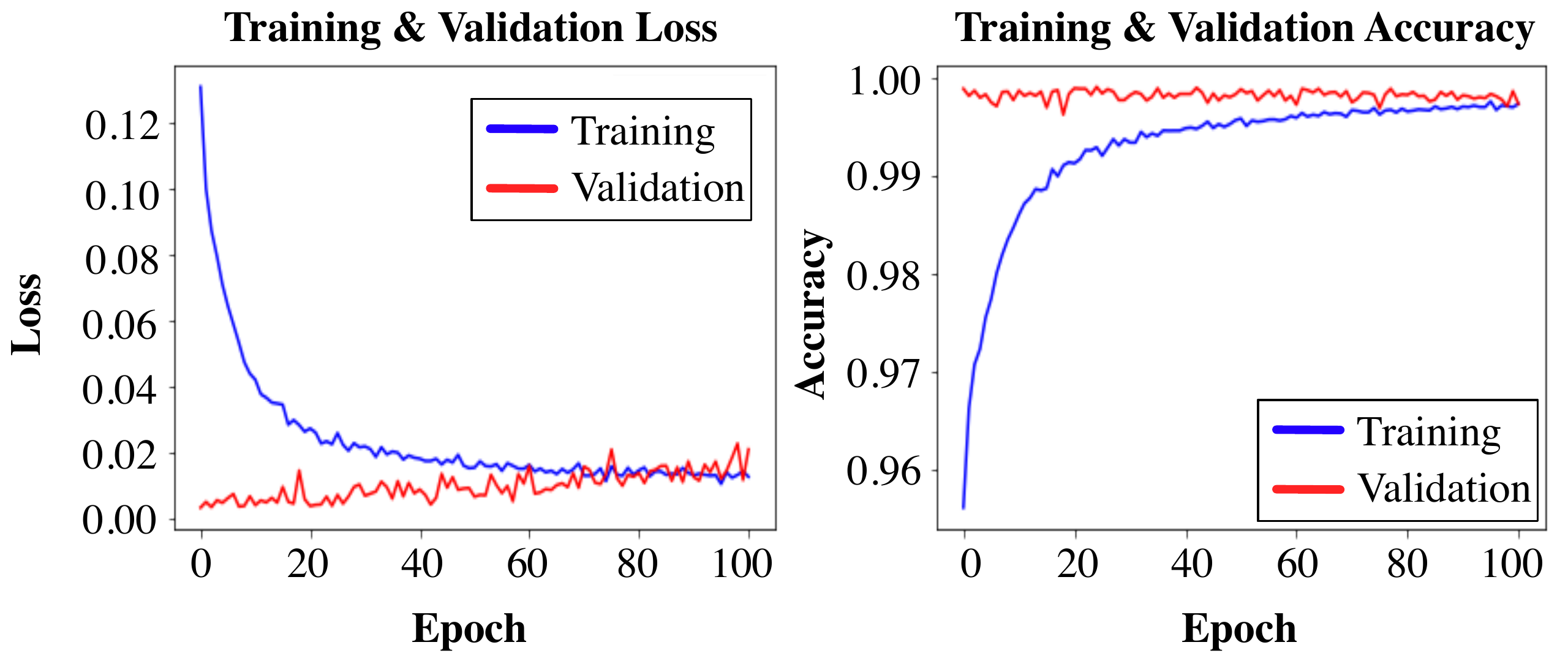}
\vspace{-0.7cm}
\caption{SecureReqNet (Shallow) Trained to Optimal Capacity. The binary cross-entropy loss is computed at each epoch. The accuracy is defined by $ Acc = (TP+TN)/(TP+TN+FP+FN)$, where $TP$ = True Positives, $TN$ = True Negatives, $FP$ = False Positives, and $FN$ = False Negatives.}
\label{fig:curves}
\vspace{0.1cm}
\end{figure}

\input{tab/results.tex}

\subsection{RQ$_1$ Results: Comparing CNN Architectures}

We present the results for all of our evaluation metrics in \tabref{tab:results}. This table provides the architecture configuration, as well as pertinent hyper-parameters including the number of kernels used in the convolutional layers, the total number of model parameters (illustrating capacity), and the number of epochs required to reach optimal network capacity and optimal performance. When comparing our measured performance metrics across models, we observe that \textit{SecureReqNet} (shallow) and $\alpha$-\textit{SecureReqNet} tend to perform best overall, with $\alpha$-\textit{SecureReqNet} achieving the best accuracy and AUC on the industrial user stories, and \textit{SecureReqNet} (shallow) achieving the best loss and accuracy on the dataset of open source issues. We also observe that \textit{Alex-SecureReqNet} performs quite poorly, which we attribute to its lack of n-gram context. One interesting note when comparing the results between models is that \textit{SecureReqNet} (deep) achieved the lowest loss on the Industry test dataset, without a large divergence from the best accuracy reported by \textit{SecureReqNet} (shallow) (0.7015 \textit{vs.} 0.7164). This indicates that while \textit{SecureReqNet} (shallow) achieved the best performance on our particular set of commercial user stories, we might expect \textit{SecureReqNet} (deep) to exhibit more stable performance on other similar datasets. \figref{fig:curves} illustrates training and validation loss and accuracy during the \textit{SecureReqNet} (shallow) training process. We observed a test validation loss of 0.1313 and accuracy of 0.9736 compared to 0.0272 and 0.9936 for the $\alpha$-\textit{SecureReqNet} respectively. Thus, we found that the shallow architecture had a better training performance through its minimization of generalization error.

\subsection{RQ$_2$ Results: Overall Performance}

When examining the overall performance of our best performing architecture configurations, we find the results to be extremely promising. On our industrial set of user stories, $\alpha$-\textit{SecureReqNet} achieved an accuracy of 0.7164. On our set of open source issues, \textit{SecureReqNet} (shallow) achieves an impressive accuracy of 0.96. The performance of \textit{SecureReqNet} (shallow) on the set of open source issues is extremely promising, as this dataset is quite diverse, and it illustrates the ability of our approach to be trained and effectively applied to unseen issues in a realistic setting.

%% file: tab/results.tex
\begin{table*}[]
\centering
\footnotesize
\vspace{-0.7cm}
\caption{Neural Network Architectures Results - Bold numbers represent best results}
\vspace{-0.2cm}
\label{tab:results}
\resizebox{\textwidth}{!}{%
\begin{tabular}{@{}lcccccccccc@{}}
\toprule
\multicolumn{1}{c}{\multirow{2}{*}{\textbf{\begin{tabular}[c]{@{}c@{}}Neural Network \\ Architecture\end{tabular}}}} & \multirow{2}{*}{\textbf{{\begin{tabular}[c]{@{}c@{}}Kernels \\ (\textit{n}-gram)\end{tabular}}}} & \multirow{2}{*}{\textbf{\#Params}} & \multirow{2}{*}{\textbf{\begin{tabular}[c]{@{}c@{}}Opt. Capacity \\ (epochs)\end{tabular}}} & \multirow{2}{*}{\textbf{\begin{tabular}[c]{@{}c@{}}Best Model\\ (epochs)\end{tabular}}} & \multicolumn{3}{c}{\textbf{\begin{tabular}[c]{@{}c@{}}Industry Test Dataset\end{tabular}}} & \multicolumn{3}{c}{\textbf{\begin{tabular}[c]{@{}c@{}}Issues Test Dataset\end{tabular}}} \\ \cmidrule(l){6-11} 
\multicolumn{1}{c}{} &  &  &  &  & \textbf{Loss} & \textbf{Acc.} & \textbf{AUC} & \textbf{Loss} & \textbf{Acc.} & \textbf{AUC} \\ \cmidrule(r){1-5}
\textit{SecureReqNet} (shallow) & [1/3/5] & 116,354 & 101 & 25 & 1.4325 & \textbf{0.7164} & 0.7 & \textbf{0.231} & \textbf{0.96} & 0.984 \\
\textit{SecureReqNet} (deep) & [1/3/5] & 662,018 & 104 & 4 & \textbf{0.859} & 0.7015 & 0.668 & 0.4044 & 0.8358 & 0.928 \\
\textit{Alex-SecureReqNet} (deep) & [7$\rightarrow$3] & 6,052,866 & 112 & 72 & 5.2794 & 0.6567 & 0.5 & 7.6972 & 0.498 & 0.5 \\
$\alpha$-\textit{SecureReqNet} (deep) & [7$\rightarrow$3] & 100,946 & 104 & 40 & 2.3793 & \textbf{0.7164} & \textbf{0.747} & 0.2615 & 0.958 & \textbf{0.986} \\ \bottomrule
\end{tabular}%
}
\vspace{-0.5cm}
\end{table*}

%% file: related-work.tex
\section{Related Work}
\label{sec:related-work}

\subsection{Automated Identification of Security Requirements}
\label{subsec:rel-reqs}

The most closely related work to the approach that we propose was conducted by Riaz \etal~\cite{Riaz:RE'14}. This approach aims to automatically identify and classify security relevant sentences in natural language requirements into one of ten objective groups for a specified domain using a \textit{k}-NN classifier and a textual distance function. In contrast to this work, our approach aims to perform a binary classification on a much larger and more diverse dataset of \textit{issues} in bug tracking systems into SR or non-SR categories. To effectively classify documents in such a diverse NL corpora our approach utilizes a two-phase DL technique that learns unsupervised word embeddings to train a CNN to classify issues. Additional related research has been conducted to identify security requirements from formal regulatory documents~\cite{Breaux:TSE'08,Maxwell:RE'11,Peclat:NGC'18}, predict severity levels with rule learning by mining defect reports  \cite{Menzis:ICSM'08}, and identify useful information from software security documents~\cite{Guo:AIRE'18}. Several studies and surveys have been conducted that examine techniques and methodologies for security requirements engineering~\cite{MelladoCSI'10,Fabian:RE'10,Howard:SDL'06,Mead:SIGSOFT'05}. 

\subsection{Applications of DL to Software Engineering Tasks}
\label{subsec:rel-dl}

Using DL to help improve automate SE tasks is becoming increasingly popular in the SE research community due to its ability to automatically identify complex features in large datasets and use these features to train effective prediction models. While a full survey of the use of DL in software engineering research is outside the scope of this paper, we review the most closely related work to our own. We adapt and build upon the work of Han \etal~\cite{Han:ICSME'18} who utilize a CNN to predict the severity of software vulnerabilities listed in the CVE database. More specifically, we adapt their CNN architecture to contain more layers, granting our network the necessary capacity to learn features from our highly diverse dataset. Chen \etal~\cite{Chen:ASE'16} utilize neural word embeddings and a CNN to help improve the proficiency of retrieving relevant results on Stack Overflow when the queries are posed in a language other than English.  Xu \etal~\cite{Xu:ASE'16} utilize a similar neural language model and CNN to link similar pieces of information in Stack Overflow posts. Gu \etal~\cite{Gu:FSE'16} use an RNN encoder-decoder model to help improve the effectiveness of searching for API call sequences using natural language queries. Guo \etal~\cite{Guo:ICSE'17} utilize neural word embeddings and an RNN to learn the semantics of existing trace links in a software project in order to make effective predictions for additional trace links. The advancements brought by these papers show the promise of DL in helping to classify and retrieve software artifacts, and motivate our approach.

%% file: conclusion.tex
\section{Conclusion \& Future Work}
\label{sec:conclusion}

In this paper we have proposed a two-stage CNN approach for learning to automatically identify security-related issues in issue tracking systems. We evaluated our approach on both a large dataset of open source issues, and a smaller dataset of industrial user stories. Our approach shows promising performance, achieving high accuracy on both datasets. In the future, we aim to perform a more in-depth empirical evaluation comparing \textit{SecureReqNet} to different baselines and apply our approach to help contextualize other software engineering tasks to security contexts. \vspace{-0.2cm}